\documentclass[journal,twoside,web]{ieeecolor}
\usepackage{generic}
\usepackage{cite}
\usepackage{amsmath,amssymb,amsfonts}
\usepackage{algorithmic}
\usepackage{graphicx}
\usepackage{algorithm,algorithmic}
\usepackage{hyperref}
\hypersetup{hidelinks=true}
\usepackage{textcomp}
\def\BibTeX{{\rm B\kern-.05em{\sc i\kern-.025em b}\kern-.08em
    T\kern-.1667em\lower.7ex\hbox{E}\kern-.125emX}}
\providecommand{\refname}{REFERENCES}
\makeatletter
\def\thebibliography#1{\section*{\color{nblue}\refname}%
    \addcontentsline{toc}{section}{\refname}%
    \footnotesize \vskip 0.3\baselineskip plus 0.1\baselineskip minus 0.1\baselineskip%
    \list{\@biblabel{\@arabic\c@enumiv}}%
    {\settowidth\labelwidth{\@biblabel{#1}}%
    \leftmargin\labelwidth
    \advance\leftmargin\labelsep\relax
    \itemsep 0pt plus 0pt\relax%
    \usecounter{enumiv}%
    \let\p@enumiv\@empty
    \renewcommand\theenumiv{\@arabic\c@enumiv}}%
    \let\@IEEElatexbibitem\bibitem%
    \def\bibitem{\@IEEEbibitemprefix\@IEEElatexbibitem}%
\def\newblock{\hskip .11em plus .33em minus .07em}%
\if@technote\sloppy\clubpenalty4000\widowpenalty4000\interlinepenalty100%
\else\sloppy\clubpenalty4000\widowpenalty4000\interlinepenalty500\fi%
    \sfcode`\.=1000\relax}
\makeatother
\begin{document}
\bstctlcite{IEEEexample:BSTcontrol}
\title{Self-Auditing Residual Drifting for Pathology-Preserving Accelerated Knee MRI}
\author{Qing Lyu, \IEEEmembership{Member, IEEE}, Jianxu Wang, Mohammad Kawas, Ge Wang, \IEEEmembership{Fellow, IEEE}, and Christopher T. Whitlow
\thanks{Q. Lyu, M. Kawas, and C.T. Whitlow are with the Department of Radiology \& Biomedical Imaging, Yale School of Medicine, New Haven, CT 06510 USA (e-mail: \{qing.lyu, mohammad.kawas, christopher.whitlow\}@yale.edu).}
\thanks{J. Wang and G. Wang are with 
the Department of Biomedical Engineering, Rensselaer Polytechnic Institute, Troy, NY 
12180 USA (e-mail: \{wangj68, wangg6\}@rpi.edu).}}

\maketitle

\begin{abstract}
Accelerated magnetic resonance imaging reduces acquisition time, but reconstruction from undersampled k-space can blur diagnostically relevant structures or introduce failures that are not captured by global image metrics. We propose SA-RDM-DC, a Self-Auditing Residual generative Drifting Model with Data Consistency for accelerated knee MRI. The method adapts the newly proposed generative drifting paradigm to accelerated MRI by training a physics-conditioned drift field from the zero-filled reconstruction toward the fully sampled residual correction. It predicts image- and missing-k-space residual corrections, enforces data consistency with acquired k-space, uses frequency-aware and residual drifting supervision to recover fine detail, and produces dense error maps and slice-level risk scores in the same inference pass. We evaluate SA-RDM-DC on multi-coil fastMRI knee data at acceleration factors of 4, 8, and 12, with fastMRI+ pathology annotations for region-level and classifier-based task preservation, and on SKM-TEA for zero-shot and fine-tuned protocol-shift evaluation. Compared with zero-filled reconstruction, UNet-image-SENSE, DC-UNet, Score-Diffusion, ELF-Diff, SENSE-VarNet, and MoDL baselines, SA-RDM-DC achieves the highest SSIM across fastMRI acceleration factors while retaining subsecond per-slice inference and avoiding the long sampling time of iterative diffusion baselines. In pathology-aware analysis, SA-RDM-DC preserves lesion-region structural fidelity and reduces meniscus prediction instability. Its self-auditing scores strongly identify high-error reconstructions on fastMRI and partially transfer as a selective-review signal under SKM-TEA protocol shift. These results support reconstruction evaluation that jointly considers image fidelity, pathology preservation, runtime, and case-specific reliability.
\end{abstract}


\section{Introduction}
\label{sec:introduction}
\IEEEPARstart{M}{agnetic} resonance imaging (MRI) provides excellent soft-tissue contrast for musculoskeletal assessment, but conventional acquisitions are time consuming \cite{lustig2007sparse,zbontar2018fastmri}. Long examinations reduce throughput and increase discomfort and motion sensitivity \cite{zbontar2018fastmri}. Accelerated MRI addresses this by acquiring fewer k-space samples and solving an inverse reconstruction problem \cite{pruessmann1999sense,griswold2002grappa,lustig2007sparse}. Classical parallel imaging and compressed sensing methods, including SENSE, GRAPPA, and sparse MRI, use coil sensitivity, calibration data, and image priors to recover missing measurements \cite{pruessmann1999sense,griswold2002grappa,lustig2007sparse}. However, high acceleration remains challenging because reconstructions must recover fine anatomy while respecting acquired measurements.

Deep learning has substantially improved accelerated MRI reconstruction by learning image priors from large datasets. Image-to-image networks, cascaded data-consistency networks, MoDL, and variational networks have demonstrated strong performance on public benchmarks such as fastMRI \cite{ronneberger2015unet,schlemper2018deepcascade,aggarwal2019modl,hammernik2018varnet,sriram2020e2evarnet,zbontar2018fastmri}. Unrolled methods such as MoDL and VarNet improve data fidelity by embedding reconstruction physics into iterative network blocks, but these approaches are still commonly optimized and compared using global metrics such as PSNR, SSIM, and NMSE. Although useful, global metrics can underweight small pathology regions, high-frequency edges, and task-relevant errors that matter in clinical interpretation \cite{zhao2022fastmriplus,shaw2021quality}.

This limitation is especially relevant in knee MRI, where meniscal tears, cartilage defects, marrow abnormalities, and ligament findings can be small, spatially localized, or dominated by high-frequency structure. A reconstruction with high global SSIM can still suppress or distort clinically important details, and a model that performs well in one acquisition setting may not preserve reliability under protocol shift. These concerns motivate evaluation beyond global fidelity, including pathology-region measurements, task-preservation analysis, and reliability assessment under acquisition-protocol changes \cite{zhao2022fastmriplus,desai2021skmtea,shaw2021quality}.

Generative reconstruction has become increasingly attractive because score-based, diffusion, and bridge-based priors can recover detailed anatomy and incorporate measurement consistency \cite{chung2022scoremri,cao2024hfs,shin2025elfdiff,cui2025spiritdiffusion,mirza2026fdb}. These methods can improve perceptual detail through iterative sampling and data consistency, but long sampling chains can be expensive for routine deployment. Drifting models offer an efficient alternative by learning residual transport through attraction-repulsion drift fields and supporting one-step inference \cite{deng2026drifting,wang2026rddm}.

Reliability estimation complements image-fidelity optimization. In MRI reconstruction, uncertainty quantification and direct error-prediction studies show that learned uncertainty or error maps can localize unreliable reconstructions and failures missed by global metrics \cite{edupuganti2021uncertainty,hu2021predict,shaw2021quality,ekanayake2025pixcue}. More broadly, medical-image quality-control and confidence-calibration studies show that automated reliability signals can triage outputs for manual review when dense ground truth is unavailable or domain shift is present \cite{robinson2019automated,mehrtash2020calibration,fournel2021misaqc,zenk2025failure}. These studies motivate self-auditing reconstruction, pairing image recovery with dense and slice-level reliability estimates.

We propose SA-RDM-DC, a self-auditing residual generative drifting model with data consistency for pathology-preserving accelerated knee MRI. SA-RDM-DC adapts the newly proposed generative drifting paradigm to the accelerated MRI inverse problem by learning a residual-domain drift field from the zero-filled reconstruction toward the fully sampled residual correction. The reconstruction subnetwork predicts image-domain and missing-k-space residual corrections, applies hard measured-k-space consistency, and uses frequency-aware and residual drifting supervision. The self-auditing subnetwork predicts dense error maps, quantile maps, and slice-level risk scores from reconstruction, residual, physics-residual, high-pass, mask, and acceleration-conditioned features. The main methodological novelty is not the isolated use of data consistency, residual learning, or uncertainty estimation, but their integration into a residual generative drifting framework: SA-RDM-DC learns drift-guided transport from the zero-filled reconstruction to the fully sampled residual correction, while hard measured-k-space projection preserves acquired data and PC-SAN estimates reconstruction reliability from the same physics-conditioned residual features.

We evaluate SA-RDM-DC on fastMRI multi-coil knee reconstruction at $R=4$, $R=8$, and $R=12$, fastMRI+ pathology-preserving analysis, and SKM-TEA zero-shot and fine-tuned protocol-shift evaluation. Our contributions are the first residual generative drifting formulation for accelerated MRI, a physics-conditioned dual-domain residual reconstruction model with hard measured-data consistency, an integrated self-auditing estimator that provides dense and slice-level reliability outputs in the same inference pass, and a benchmark that combines global metrics, pathology-region fidelity, task-preservation, runtime, and protocol-shift reliability.

\section{Materials and Methodology}
\label{sec:materials_methods}

\subsection{Problem Formulation}
\label{subsec:problem_formulation}
Let $x \in \mathbb{C}^{H \times W}$ denote the fully sampled complex-valued target image, $S_c$ the sensitivity map for coil $c$, $\mathcal{F}$ the two-dimensional Fourier transform, and $M \in \{0,1\}^{H_k \times W_k}$ the undersampling mask. We define the unmasked multi-coil forward operator and its masked counterpart as
\begin{equation}
    \mathcal{A}x=\{\mathcal{F}(S_c x)\}_{c=1}^{C},
    \quad
    \mathcal{A}_M x = M \odot \mathcal{A}x .
    \label{eq:forward_operators}
\end{equation}
The multi-coil accelerated MRI forward model is
\begin{equation}
    y = \mathcal{A}_M x + \eta,
    \label{eq:forward_model}
\end{equation}
where $y=\{y_c\}_{c=1}^{C}$ is the acquired undersampled multi-coil k-space, $\eta=\{\eta_c\}_{c=1}^{C}$ denotes measurement noise, and $\odot$ is elementwise multiplication with mask broadcasting across coils. Reconstruction seeks an estimate $\hat{x}$ from $y$:
\begin{equation}
    \hat{x} = \mathcal{R}_{\theta}(y,M,S,R),
    \label{eq:reconstruction_mapping}
\end{equation}
where $S=\{S_c\}_{c=1}^{C}$ and $R$ is the nominal acceleration factor. The zero-filled SENSE reconstruction is used as the image-domain input,
\begin{equation}
    x_{\mathrm{zf}} = \mathcal{A}^{H}(y),
    \label{eq:zero_filled}
\end{equation}
where $\mathcal{A}^{H}$ denotes the coil-combined adjoint reconstruction after inserting zeros at unmeasured k-space locations. The goal is not only to minimize image error, but also to preserve pathology-relevant detail and estimate a reliability map $\hat{e}$ that predicts the absolute reconstruction error.

\begin{figure*}[!t]
\centerline{\includegraphics[width=0.9\textwidth]{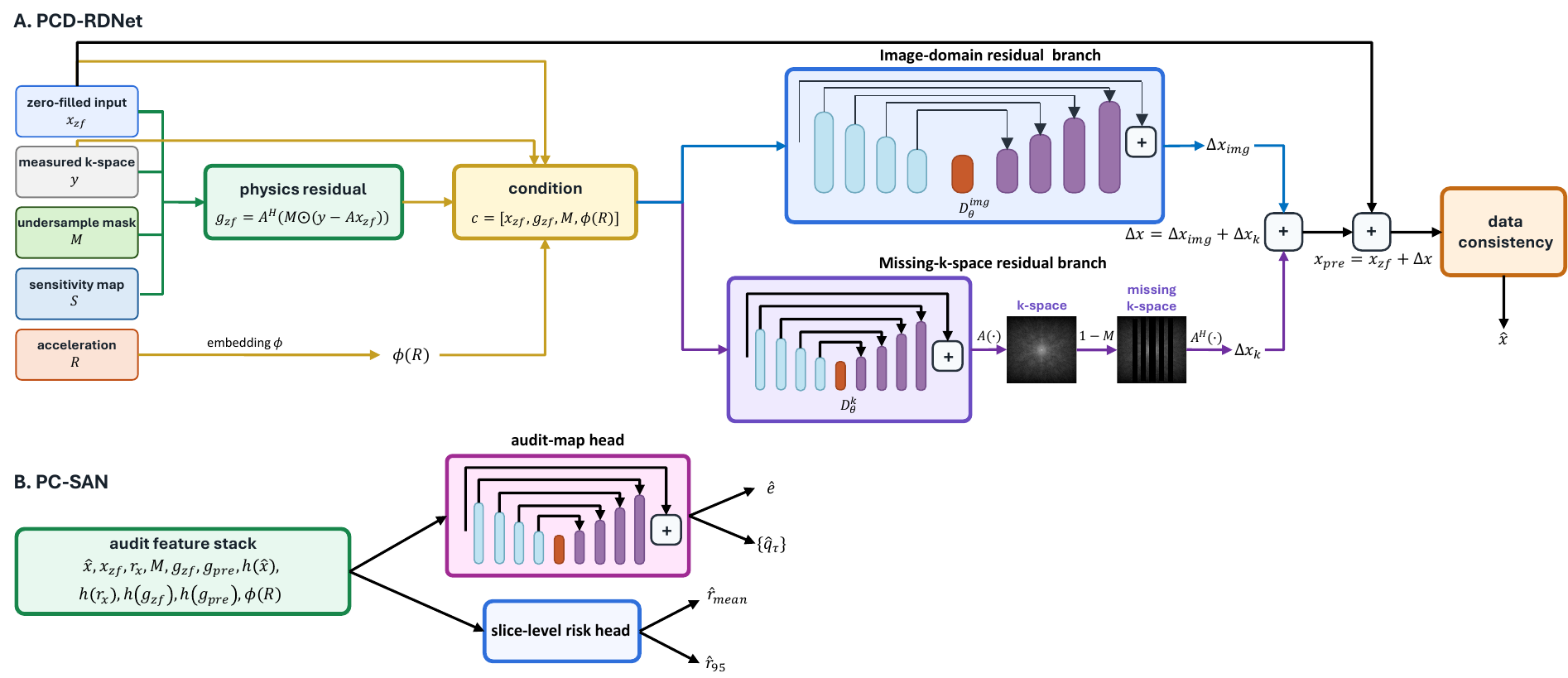}}
\caption{Overview of SA-RDM-DC. A: PCD-RDNet predicts image-domain and missing-k-space residuals from physics-conditioned inputs and applies measured-data consistency. B: PC-SAN predicts error maps, quantile maps, and slice-level risk scores from reconstruction and physics-residual features.}
\label{fig:sa_rdm_dc_framework}
\end{figure*}

\subsection{Datasets and Preprocessing}
\label{subsec:datasets_preprocessing}

\subsubsection{fastMRI and fastMRI+}
The fastMRI dataset is a public raw-data benchmark for accelerated MRI reconstruction \cite{zbontar2018fastmri}. The fastMRI knee k-space dataset contains fully sampled coronal proton-density-weighted acquisitions with and without frequency-selective fat suppression, acquired using Cartesian two-dimensional turbo spin-echo sequences on clinical Siemens systems, including 3-T Magnetom Skyra, Prisma, and Biograph-mMR scanners and a 1.5-T Magnetom Aera scanner. The original fastMRI knee acquisitions have approximately submillimeter in-plane resolution and 3-mm slice thickness. fastMRI+ extends fastMRI with clinical pathology annotations for knee and brain MRI \cite{zhao2022fastmriplus}. In fastMRI+, board-certified radiologists reviewed image quality and performed clinical annotation. Pathologies were labeled on a slice-by-slice basis using bounding boxes associated with predefined disease categories, and study-level labels were also provided. The released fastMRI+ knee annotations include 16,154 bounding boxes and 13 study-level labels across 22 pathology categories.

\subsubsection{SKM-TEA}
SKM-TEA is a public quantitative knee MRI dataset designed for multi-task evaluation of reconstruction and image analysis methods \cite{desai2021skmtea}. It contains 155 anonymized patient scans, approximately 25,000 slices, raw multi-coil k-space, scanner-generated DICOM images, SENSE reconstructions, tissue segmentations, and bounding-box annotations for 16 clinically relevant pathologies. The dataset provides DICOM-track and raw-track masks for segmentation and quantitative-biomarker evaluation on images reconstructed from k-space. The scans were acquired at Stanford Health Care using sagittal three-dimensional quantitative double-echo steady-state (qDESS) imaging on two 3-T GE MR750 scanners. Each qDESS acquisition contains two inherently registered echoes. The acquisition protocol used 2$\times$1 parallel imaging with elliptical sampling, a $416 \times 512$ readout-by-phase-encode matrix, $0.38 \times 0.31$ mm$^2$ in-plane resolution, and 80--88 acquired slices depending on knee size.

\subsubsection{Data Preprocessing and Undersampling}
For the fastMRI experiments, 973 multi-coil knee MRI scans were selected, including 484 proton-density-weighted volumes and 489 fat-suppressed proton-density-weighted volumes. Volumes were split at the scan level before slice extraction to avoid leakage between training and validation; 850 scans were assigned to training and 123 to validation and testing. Original multi-coil data were coil-compressed to 12 virtual coils before reconstruction experiments. Images were center-cropped to $320 \times 320$ and normalized by a robust volume-level intensity scale. Coil sensitivity maps were estimated from low-frequency calibration data and saved with the preprocessed volumes. Among the 973 processed fastMRI volumes, 815 overlapped with reviewed fastMRI+ knee annotations, providing 13,653 pathology bounding boxes including meniscus, cartilage, and other categories for pathology-aware analysis. The preprocessing pipeline stored the reference SENSE image, zero-filled SENSE image, undersampled k-space, sampling mask, and sensitivity maps for each acceleration. For SKM-TEA, 119 scans were used for model fine-tuning and 36 scans for validation and evaluation. The raw-data track used hybrid k-space and SENSE targets from the qDESS acquisition; two-dimensional slices were reconstructed in phase-encode planes, and echo 1 was used to focus the protocol-shift and task-preservation analyses. The SKM-TEA experiment was designed as a protocol-shift analysis with two settings: zero-shot transfer, which tests direct generalization from fastMRI to a different acquisition protocol, and target-protocol fine-tuning, which tests adaptability after exposure to SKM-TEA training data. Random masks were generated with center fractions of 0.08 for $R=4$ and 0.04 for higher accelerations. The main experiments used acceleration factors $R=4$, $R=8$, and $R=12$, with $R=12$ treated as a high-acceleration stress condition.

\subsection{SA-RDM-DC}
\label{subsec:sa_rdm_dc}

SA-RDM-DC consists of two coupled subnetworks. The reconstruction subnetwork, termed PCD-RDNet, performs physics-conditioned dual-domain residual drifting from the zero-filled SENSE image to a data-consistent reconstruction. The self-auditing subnetwork, termed PC-SAN, uses image-domain and measured-data residual features to predict dense error maps, quantile error maps, and slice-level risk scores. During inference, SA-RDM-DC produces both the reconstruction and audit outputs in a single forward pass. During training, PCD-RDNet is optimized first with reconstruction and conditional drifting losses; PC-SAN is then trained with reconstruction parameters frozen so that audit calibration does not degrade image quality.

The overall architecture is illustrated in Fig.~\ref{fig:sa_rdm_dc_framework}. The following subsections describe the reconstruction and self-auditing subnetworks corresponding to panels A and B, respectively.

\subsubsection{PCD-RDNet: Physics-Conditioned Dual-Domain Residual Drifting Network}
PCD-RDNet contains four key components: physics-conditioned inputs, an image-domain residual branch, a missing-k-space residual branch, and hard measured-data consistency. In addition to the zero-filled image $x_{\mathrm{zf}}$, the network receives the sampling mask $M$, a learned acceleration embedding $\phi(R)$, and a physics residual $g_{\mathrm{zf}}$ computed from the measured k-space mismatch of the zero-filled input:
\begin{equation}
    g_{\mathrm{zf}} =
    \mathcal{A}^{H}\left(M\odot\left(y-\mathcal{A}x_{\mathrm{zf}}\right)\right).
    \label{eq:zf_physics_residual}
\end{equation}
The reconstruction condition is the channel-wise concatenation
\begin{equation}
    c_{\mathrm{rec}} =
    x_{\mathrm{zf}}\oplus M \oplus g_{\mathrm{zf}}\oplus \phi(R).
    \label{eq:recon_condition}
\end{equation}
PCD-RDNet uses two branches with residual attention U-Net backbones. The image branch predicts an image-domain residual,
\begin{equation}
    \Delta x_{\mathrm{img}} = D_{\theta}^{\mathrm{img}}\left(c_{\mathrm{rec}}\right),
    \label{eq:image_residual_branch}
\end{equation}
and the missing-k-space branch predicts an image-shaped correction that is projected through unmeasured k-space:
\begin{equation}
    \Delta x_{\mathrm{k}} =
    \mathcal{A}^{H}\left((1-M)\odot
    \mathcal{A}D_{\theta}^{\mathrm{k}}\left(c_{\mathrm{rec}}\right)\right).
    \label{eq:kspace_residual_branch}
\end{equation}
The pre-consistency reconstruction is therefore
\begin{equation}
    x_{\mathrm{pre}} =
    x_{\mathrm{zf}}+\Delta x_{\mathrm{img}}+\Delta x_{\mathrm{k}}.
    \label{eq:pre_dc_reconstruction}
\end{equation}
Data consistency is then applied by replacing acquired k-space samples with the measured data:
\begin{equation}
    \hat{x} =
    \mathcal{A}^{H}\left(M\odot y+(1-M)\odot \mathcal{A}x_{\mathrm{pre}}\right).
    \label{eq:hard_dc_image}
\end{equation}
Because sensitivity maps are estimated from calibration data, this projection enforces consistency with the implemented SENSE forward/adjoint model by copying measured entries before final coil-combined reconstruction. This one-step dual-domain design keeps inference feed-forward while allowing the learned update to focus on recovering missing frequencies while preserving acquired measurements.

PCD-RDNet is trained with a base reconstruction objective and a conditional drifting objective. The base reconstruction objective is
\begin{equation}
\begin{split}
    \mathcal{L}_{\mathrm{base}} =
    \lambda_{1}\mathcal{L}_{1}
    + \lambda_{\mathrm{mse}}\mathcal{L}_{\mathrm{mse}}
    + \lambda_{\mathrm{ssim}}\mathcal{L}_{\mathrm{ssim}} \\
    + \lambda_{\mathrm{freq}}\mathcal{L}_{\mathrm{freq}}
    + \lambda_{\mathrm{dc}}\mathcal{L}_{\mathrm{dc}},
\end{split}
\label{eq:base_recon_objective}
\end{equation}
where
\begin{equation}
    \mathcal{L}_{1} = \left\|m(\hat{x}) - m(x)\right\|_{1},
    \quad
    \mathcal{L}_{\mathrm{mse}} = \left\|\hat{x}-x\right\|_{2}^{2},
    \label{eq:image_fidelity_losses}
\end{equation}
\begin{equation}
    \mathcal{L}_{\mathrm{ssim}} =
    1-\mathrm{SSIM}\left(m(\hat{x}),m(x)\right),
    \label{eq:ssim_loss}
\end{equation}
and SSIM is the structural similarity index \cite{wang2004image}. The missing-k-space and pre-consistency data-consistency losses are
\begin{equation}
    \mathcal{L}_{\mathrm{freq}} =
    \left\|(1-M)\odot \left(\mathcal{A}\hat{x}-\mathcal{A}x\right)\right\|_{2}^{2},
    \label{eq:freq_loss}
\end{equation}
\begin{equation}
    \mathcal{L}_{\mathrm{dc}} =
    \left\|\mathcal{A}_{M}x_{\mathrm{pre}}-y\right\|_{2}^{2}.
    \label{eq:dc_loss}
\end{equation}
The base objective combines complementary constraints: $\mathcal{L}_{1}$ and $\mathcal{L}_{mse}$ enforce voxel-wise image fidelity, $\mathcal{L}_{ssim}$ promotes structural agreement in normalized magnitude images, $\mathcal{L}_{\mathrm{freq}}$ directly supervises unacquired k-space recovery, and $\mathcal{L}_{\mathrm{dc}}$ discourages learned residual updates from conflicting with acquired measurements before hard data consistency. Together, these terms balance image-domain accuracy, missing-frequency recovery, and measurement fidelity.

For residual generative drifting supervision, the generated residual $r_{\theta}=\hat{x}-x_{\mathrm{zf}}$ is attracted toward the fully sampled residual $r^{+}=x-x_{\mathrm{zf}}$ and repelled from fixed negative residuals such as the zero-filled residual $r^{-}=0$. A conditional drift field $V(r_{\theta};c)$ is computed using learned residual features together with acceleration and mask conditioning; this field defines the direction of the residual-domain generative transport used during training. The stop-gradient drifting target is projected through hard data consistency:
\begin{equation}
    x_{\mathrm{drift}} =
    \Pi_{y}\left(x_{\mathrm{zf}}+r_{\theta}+\gamma V(r_{\theta};c)\right),
    \label{eq:dc_projected_drift_target}
\end{equation}
where $\Pi_{y}(\cdot)$ denotes the hard measured-k-space projection in Eq. (\ref{eq:hard_dc_image}). The drifting loss is
\begin{equation}
    \mathcal{L}_{\mathrm{drift}} =
    \left\|\hat{x}-\operatorname{sg}\left(x_{\mathrm{drift}}\right)\right\|_{2}^{2}
    + \lambda_{\mathrm{feat}}\mathcal{L}_{\mathrm{feat}},
    \label{eq:drift_loss}
\end{equation}
where $\operatorname{sg}(\cdot)$ stops gradients through the drifting target and $\mathcal{L}_{\mathrm{feat}}$ is an auxiliary contrastive loss that aligns generated residual features with paired target residual features while separating them from zero-filled residual features. The first-stage objective is
\begin{equation}
    \mathcal{L}_{\mathrm{rec}} =
    \mathcal{L}_{\mathrm{base}}+\lambda_{\mathrm{drift}}(t)\mathcal{L}_{\mathrm{drift}},
    \label{eq:recon_objective}
\end{equation}
where $\lambda_{\mathrm{drift}}(t)$ is warmed up and ramped during the reconstruction phase.

\subsubsection{PC-SAN: Physics-Calibrated Self-Auditing Network}
PC-SAN predicts reconstruction reliability from the PCD-RDNet output $\hat{x}$ and its associated image- and physics-residual features. It contains two audit output branches. A UNet-style dense audit-map head~\cite{ronneberger2015unet} predicts the pixel-wise error map $\hat{e}$ and quantile error maps $\{\hat{q}_{\tau}\}$ ($\tau\in\{0.5,0.9,0.95\}$) for calibrated pixel-wise upper bounds on reconstruction error. In parallel, a separate trainable slice-level risk head processes the same audit feature stack using residual squeeze-and-excitation blocks~\cite{hu2018squeeze}, global average pooling, and a linear output layer to predict scalar mean-risk and tail-risk scores. In addition to $x_{\mathrm{zf}}$, $\hat{x}$, their residual $\hat{x}-x_{\mathrm{zf}}$, the mask, and the acceleration embedding, PC-SAN uses the zero-filled physics residual $g_{\mathrm{zf}}$ and the pre-consistency residual
\begin{equation}
    g_{\mathrm{pre}} =
    \mathcal{A}^{H}\left(M\odot\left(y-\mathcal{A}x_{\mathrm{pre}}\right)\right).
    \label{eq:pre_dc_physics_residual}
\end{equation}
High-pass magnitude maps of $\hat{x}$, $\hat{x}-x_{\mathrm{zf}}$, $g_{\mathrm{zf}}$, and $g_{\mathrm{pre}}$ are also included as audit features. Let $r_x=\hat{x}-x_{\mathrm{zf}}$. The audit feature stack can be written compactly as
\begin{equation}
\begin{split}
    c_{\mathrm{audit}} =
    \hat{x}\oplus x_{\mathrm{zf}}\oplus r_x
    \oplus M\oplus g_{\mathrm{zf}}\oplus g_{\mathrm{pre}} \\
    \oplus h(\hat{x})\oplus h(r_x)\oplus h(g_{\mathrm{zf}}) \\
    \oplus h(g_{\mathrm{pre}})\oplus \phi(R),
\end{split}
\label{eq:audit_condition}
\end{equation}
where $h(\cdot)$ denotes the fixed high-pass magnitude feature operator, implemented as the normalized magnitude image minus a $5\times5$ average-pooled version of the same image. 

The slice-level risk scores are learned scalar outputs. Let $H_{\psi}$ denote the trainable risk head. It maps the audit feature stack to two raw scalar outputs:
\begin{equation}
    [a_{\mathrm{mean}},a_{95}] = H_{\psi}(c_{\mathrm{audit}}).
    \label{eq:risk_head_raw}
\end{equation}
The predicted slice-level mean-risk and tail-risk scores are
\begin{equation}
\begin{aligned}
    \hat{r}_{\mathrm{mean}}
    &= s\,\operatorname{softplus}(a_{\mathrm{mean}})+\epsilon, \\
    \hat{r}_{95}
    &= \hat{r}_{\mathrm{mean}}
    + s\,\operatorname{softplus}(a_{95})+\epsilon,
\end{aligned}
\label{eq:audit_scores}
\end{equation}
where $s$ is the audit error scale and $\epsilon$ is a small positivity floor. This parameterization keeps the predicted tail-risk score no smaller than the predicted mean-risk score. The mean-risk score estimates the expected overall reconstruction error, whereas the 95th-percentile score estimates localized high-error regions that may be clinically important.

PC-SAN is optimized with a composite self-auditing loss that supervises dense error prediction, slice-level risk estimation, and calibrated quantile error maps. The PC-SAN training target is the pixel-wise absolute reconstruction error
\begin{equation}
    e^{\ast} = \left|m(\hat{x})-m(x)\right|,
    \label{eq:error_target}
\end{equation}
where $m(\cdot)$ denotes the normalized magnitude mapping. The dense audit-map head is supervised by a weighted $\mathcal{L}_1$ loss,
\begin{equation}
    \mathcal{L}_{\mathrm{map}} =
    \frac{1}{N}\sum_{p}
    \left(1+\alpha\mathbb{I}\left[e^{\ast}_{p}\ge Q_{\rho}(e^{\ast})\right]\right)
    \left|\hat{e}_{p}-e^{\ast}_{p}\right|,
    \label{eq:weighted_audit_map_loss}
\end{equation}
where $Q_{\rho}$ is the per-slice $\rho$-quantile of the true error map and the weighting term emphasizes high-error pixels.

The corresponding true slice-level summaries are computed from $e^{\ast}$ over the image support $\Omega$:
\begin{equation}
    r^{\ast}_{\mathrm{mean}} =
    \frac{1}{|\Omega|}\sum_{p\in\Omega}e^{\ast}_{p},
    \quad
    r^{\ast}_{95} =
    Q_{0.95}\left(\{e^{\ast}_{p}:p\in\Omega\}\right),
    \label{eq:true_audit_scores}
\end{equation}
where $Q_{0.95}(\cdot)$ denotes the empirical 95th percentile operator. The slice-level risk head is trained to regress these true error summaries:
\begin{equation}
    \mathcal{L}_{\mathrm{risk}} =
    \mathrm{SmoothL1}\left(\hat{r}_{\mathrm{mean}},r^{\ast}_{\mathrm{mean}}\right)
    +
    \mathrm{SmoothL1}\left(\hat{r}_{95},r^{\ast}_{95}\right),
    \label{eq:risk_loss}
\end{equation}
where $\mathrm{SmoothL1}(\cdot,\cdot)$ denotes the Huber-style smooth $\mathcal{L}_1$ loss~\cite{huber1964robust}, which behaves quadratically for small residuals and linearly for large residuals.

For the quantile error maps, PC-SAN uses a pinball quantile-regression loss~\cite{koenker1978regression} over $\mathcal{T}=\{0.5,0.9,0.95\}$:
\begin{equation}
\begin{aligned}
    \mathcal{L}_{\mathrm{quant}} =
    \frac{1}{|\mathcal{T}|}\sum_{\tau\in\mathcal{T}}
    \frac{1}{N}\sum_{p}
    \max\left(\tau d_{\tau,p},(\tau-1)d_{\tau,p}\right), \\
    d_{\tau,p}=e^{\ast}_{p}-\hat{q}_{\tau,p}.
\end{aligned}
\label{eq:quantile_audit_loss}
\end{equation}
This loss encourages $\hat{q}_{\tau,p}$ to estimate the $\tau$-level upper quantile of the pixel-wise reconstruction error.

The complete PC-SAN objective is
\begin{equation}
    \mathcal{L}_{\mathrm{audit}} =
    \lambda_{\mathrm{map}}\mathcal{L}_{\mathrm{map}}
    + \lambda_{\mathrm{risk}}\mathcal{L}_{\mathrm{risk}}
    + \lambda_{\mathrm{quant}}\mathcal{L}_{\mathrm{quant}} .
    \label{eq:audit_loss}
\end{equation}

\subsection{Evaluation Protocol}
\label{subsec:evaluation_protocol}

\subsubsection{Global Reconstruction Evaluation}
Global reconstruction quality was measured using PSNR, SSIM, NMSE, MAE, and high-frequency error norm (HFEN). All reported image-domain metrics were computed on normalized magnitude images unless otherwise specified. HFEN was computed as the norm of the difference after Laplacian-of-Gaussian filtering. Runtime per slice and the number of model parameters were recorded to compare feed-forward, unrolled, and diffusion-based methods.

\subsubsection{Pathology-Aware Evaluation}
Pathology preservation was evaluated using fastMRI+ bounding boxes. For each annotated box $B$, metrics were computed based on the provided pathology box. ROI-level SSIM, HFEN, and edge error were computed. Edge preservation inside pathology boxes was measured by Sobel-gradient magnitude error:
\begin{equation}
    E_{\mathrm{edge}}(B) =
    \frac{1}{|B|}\sum_{p\in B}
    \left|\left|\nabla \hat{u}_{p}\right|-\left|\nabla u_{p}\right|\right|.
    \label{eq:edge_error}
\end{equation}
Box-level metrics were first averaged within each scan and then summarized across scans so that scans with many boxes did not dominate the evaluation.

\subsubsection{Task-Preservation Evaluation}
A frozen scan-level multi-label pathology classifier was used to assess whether reconstructions preserve downstream clinical predictions. The classifier was trained only on fully sampled reference SENSE images from fastMRI+ and was not exposed to accelerated or reconstructed images during training. It used a multiple-instance learning design: each scan was represented as a stack of normalized magnitude slices, each slice was encoded by a lightweight 2D convolutional encoder, and slice features were aggregated by attention pooling~\cite{ilse2018attention} into a scan-level feature vector. A linear multi-label prediction head produced sigmoid probabilities for three endpoints: any pathology, meniscal abnormality, and cartilage abnormality.

Let $u$ and $\hat{u}$ denote the reference and reconstructed scan-level image stacks, respectively. For each endpoint, the frozen classifier $C_{\omega}$ produced reference and reconstruction probabilities
\begin{equation}
    p_{\mathrm{ref}} = C_{\omega}(u),\quad
    p_{\mathrm{rec}} = C_{\omega}(\hat{u}),
    \label{eq:classifier_probs}
\end{equation}
where the classifier weights $\omega$ were fixed during reconstruction evaluation. With a validation-selected decision threshold $\tau$, prediction flips were defined as
\begin{equation}
    f_{\mathrm{flip}} =
    \mathbb{I}\left[\mathbb{I}(p_{\mathrm{rec}}\ge \tau)
    \ne \mathbb{I}(p_{\mathrm{ref}}\ge \tau)\right],
    \label{eq:classifier_flip}
\end{equation}
where $\mathbb{I}[\cdot]$ denotes the indicator function, which equals 1 when its argument is true and 0 otherwise.

\begin{figure*}[!t]
\centerline{\includegraphics[width=0.96\textwidth]{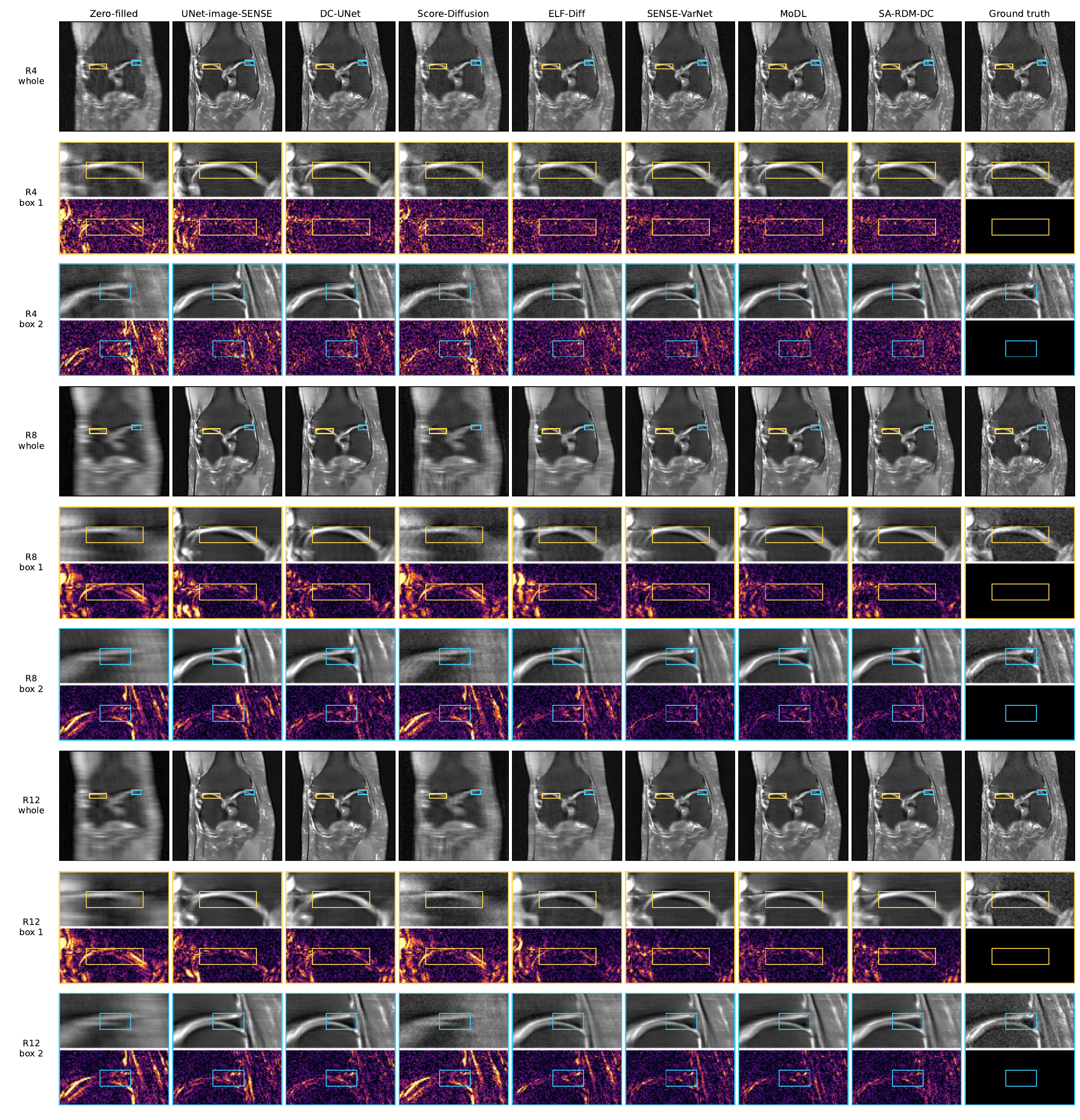}}
\caption{Qualitative reconstruction comparison on fastMRI knee data at $R=4$, $R=8$, and $R=12$. SA-RDM-DC is compared with zero-filled reconstruction, UNet-image-SENSE, DC-UNet, Score-Diffusion, ELF-Diff, SENSE-VarNet, and MoDL. Yellow and blue boxes denote fastMRI+ pathology annotations for cartilage partial-thickness loss and meniscus tear, respectively.}
\label{fig:acceleration_qualitative}
\end{figure*}

\subsubsection{Self-Auditing Reliability Evaluation}
Self-auditing was evaluated by comparing $\hat{e}$ with true absolute error and by testing whether audit-derived risk scores detect reconstruction failures. Slice-level correlation was measured between $\hat{r}_{\mathrm{mean}}$ or $\hat{r}_{95}$ and the corresponding true error summaries. Image failure was defined using low SSIM, high NMSE, or their union:
\begin{equation}
    f_{\mathrm{img}} =
    \mathbb{I}\left[\mathrm{SSIM}\le \tau_{\mathrm{ssim}}\ \mathrm{or}\ 
    \mathrm{NMSE}\ge \tau_{\mathrm{nmse}}\right].
    \label{eq:image_failure}
\end{equation}
Audit scores were evaluated using AUROC, AUPRC, and risk-coverage curves~\cite{geifman2017selective}. A simple residual baseline, $|m(\hat{x})-m(x_{\mathrm{zf}})|$, was included to test whether the learned audit head captures reconstruction risk beyond the magnitude of the network correction.

\begin{table*}[!t]
\caption{Global reconstruction comparison on fastMRI knee validation data. mean$\pm$std.}
\label{tab:global_comparison}
\centering
\footnotesize
\setlength{\tabcolsep}{2pt}
\begin{tabular*}{\textwidth}{@{\extracolsep{\fill}}lccc | ccc | ccc}
\hline
 & \multicolumn{3}{c}{$R=4$} & \multicolumn{3}{c}{$R=8$} & \multicolumn{3}{c}{$R=12$} \\
\cline{2-10}
Method & SSIM & PSNR & NMSE & SSIM & PSNR & NMSE & SSIM & PSNR & NMSE \\
\hline
Zero-filled & 0.753$\pm$0.049 & 28.63$\pm$1.37 & 0.027$\pm$0.010 & 0.614$\pm$0.061 & 24.93$\pm$1.63 & 0.061$\pm$0.017 & 0.589$\pm$0.064 & 24.59$\pm$1.62 & 0.066$\pm$0.018 \\
UNet-image-SENSE & 0.840$\pm$0.069 & 32.79$\pm$1.99 & 0.012$\pm$0.009 & 0.786$\pm$0.084 & 30.59$\pm$1.57 & 0.019$\pm$0.011 & 0.757$\pm$0.088 & 29.76$\pm$1.53 & 0.022$\pm$0.012 \\
DC-UNet & 0.852$\pm$0.071 & 33.32$\pm$2.20 & 0.011$\pm$0.009 & 0.792$\pm$0.082 & 30.91$\pm$1.63 & 0.018$\pm$0.011 & 0.763$\pm$0.087 & 29.94$\pm$1.55 & 0.022$\pm$0.012 \\
Score-Diffusion & 0.831$\pm$0.074 & 32.66$\pm$2.35 & 0.015$\pm$0.011 & 0.694$\pm$0.088 & 26.81$\pm$1.94 & 0.054$\pm$0.013 & 0.609$\pm$0.091 & 26.08$\pm$1.75 & 0.059$\pm$0.013 \\
ELF-Diff & 0.832$\pm$0.072 & 32.62$\pm$2.33 & 0.015$\pm$0.010 & 0.758$\pm$0.085 & 29.01$\pm$1.87 & 0.031$\pm$0.011 & 0.726$\pm$0.087 & 28.14$\pm$1.71 & 0.034$\pm$0.012 \\
SENSE-VarNet & 0.862$\pm$0.074 & 34.23$\pm$2.70 & 0.010$\pm$0.009 & 0.807$\pm$0.086 & 31.79$\pm$1.96 & 0.015$\pm$0.011 & 0.777$\pm$0.089 & 30.68$\pm$1.80 & 0.019$\pm$0.012 \\
MoDL & 0.861$\pm$0.076 & 34.40$\pm$2.83 & 0.010$\pm$0.009 & 0.811$\pm$0.086 & 32.13$\pm$2.09 & 0.016$\pm$0.011 & 0.782$\pm$0.088 & 30.94$\pm$1.89 & 0.018$\pm$0.012 \\
SA-RDM-DC & 0.866$\pm$0.073 & 34.29$\pm$2.54 & 0.010$\pm$0.009 & 0.814$\pm$0.082 & 31.77$\pm$2.02 & 0.015$\pm$0.012 & 0.786$\pm$0.086 & 30.74$\pm$1.83 & 0.018$\pm$0.012 \\
\hline
\end{tabular*}
\end{table*}

\subsection{Experimental Details}
\label{subsec:experimental_details}
SA-RDM-DC was compared with zero-filled reconstruction, UNet-image-SENSE~\cite{ronneberger2015unet}, DC-UNet~\cite{schlemper2018deepcascade}, SENSE-VarNet~\cite{hammernik2018varnet,sriram2020e2evarnet}, MoDL~\cite{aggarwal2019modl}, Score-Diffusion~\cite{chung2022scoremri}, and ELF-Diff~\cite{shin2025elfdiff} on matched masks and scan splits. VarNet used six cascades, and MoDL used five unrolled reconstruction iterations. Score-Diffusion used 200 sampling steps with one corrector step per predictor step, while ELF-Diff used 50 respaced reverse-diffusion steps with ensemble size 8. 

SA-RDM-DC training used AdamW optimization~\cite{loshchilov2019decoupled} with learning rate $10^{-4}$, weight decay $10^{-5}$, mixed precision when available, and scan-level validation after each epoch. The PCD-RDNet was trained first and then frozen for PC-SAN training. In the default two-stage configuration, PCD-RDNet training was set for 70 epochs with early stopping patience of 10 epochs, and PC-SAN training continued another 50 epochs with early stopping patience of 10 epochs. The audit objective weight was linearly ramped during the first audit epochs to avoid unstable calibration. Target-protocol fine-tuning on SKM-TEA was conducted with learning rate $2\times10^{-5}$ and continued for 30 epochs with early stopping patience of 7 epochs. All experiments were conducted using PyTorch~\cite{paszke2019pytorch} on a single NVIDIA RTX PRO 6000 Blackwell GPU with 96 GB memory.

\begin{table*}[!t]
\caption{Pathology-region fidelity and classifier task-preservation metrics on fastMRI+.}
\label{tab:pathology_results}
\centering
\footnotesize
\setlength{\tabcolsep}{2pt}
\begin{tabular*}{\textwidth}{@{\extracolsep{\fill}}llcccccccc}
\hline
Accel. & Metric & Zero-filled & UNet-image-SENSE & DC-UNet & Score-Diffusion & ELF-Diff & SENSE-VarNet & MoDL & SA-RDM-DC \\
\hline
$R=4$ & ROI SSIM & 0.861 & 0.924 & 0.934 & 0.908 & 0.924 & 0.945 & 0.945 & 0.948 \\
$R=4$ & HFEN & 0.584 & 0.430 & 0.389 & 0.450 & 0.426 & 0.341 & 0.329 & 0.325 \\
$R=4$ & Edge error & 0.026 & 0.020 & 0.019 & 0.022 & 0.021 & 0.0178 & 0.018 & 0.017 \\
$R=4$ & Meniscus flip & 0.061 & 0.102 & 0.082 & 0.102 & 0.092 & 0.092 & 0.112 & 0.051 \\
$R=4$ & Cartilage flip & 0.122 & 0.092 & 0.092 & 0.122 & 0.102 & 0.092 & 0.092 & 0.122 \\
\hline
$R=8$ & ROI SSIM & 0.685 & 0.861 & 0.877 & 0.771 & 0.829 & 0.898 & 0.909 & 0.910 \\
$R=8$ & HFEN & 0.756 & 0.591 & 0.542 & 0.688 & 0.635 & 0.494 & 0.456 & 0.463 \\
$R=8$ & Edge error & 0.034 & 0.026 & 0.025 & 0.030 & 0.028 & 0.023 & 0.022 & 0.023 \\
$R=8$ & Meniscus flip & 0.082 & 0.122 & 0.092 & 0.122 & 0.112 & 0.112 & 0.122 & 0.071 \\
$R=8$ & Cartilage flip & 0.153 & 0.102 & 0.112 & 0.153 & 0.132 & 0.112 & 0.102 & 0.092 \\
\hline
$R=12$ & ROI SSIM & 0.666 & 0.842 & 0.851 & 0.741 & 0.803 & 0.871 & 0.874 & 0.877 \\
$R=12$ & HFEN & 0.783 & 0.649 & 0.623 & 0.729 & 0.685 & 0.589 & 0.586 & 0.582 \\
$R=12$ & Edge error & 0.035 & 0.029 & 0.027 & 0.031 & 0.030 & 0.026 & 0.025 & 0.026 \\
$R=12$ & Meniscus flip & 0.092 & 0.133 & 0.143 & 0.153 & 0.112 & 0.102 & 0.133 & 0.102 \\
$R=12$ & Cartilage flip & 0.194 & 0.133 & 0.122 & 0.153 & 0.133 & 0.102 & 0.102 & 0.112 \\
\hline
\end{tabular*}
\end{table*}

\section{Results}
\label{sec:results}

\subsection{Global Reconstruction Performance Under MRI Acceleration}
\label{subsec:results_acceleration}
SA-RDM-DC MRI acceleration results were evaluated against zero-filled reconstruction, UNet-image-SENSE, DC-UNet, Score-Diffusion, ELF-Diff, SENSE-VarNet, and MoDL on fastMRI knee data at $R=4$, $R=8$, and $R=12$ (Table~\ref{tab:global_comparison}). SA-RDM-DC achieved the highest SSIM at all accelerations: 0.866, 0.814, and 0.786, respectively. It was competitive with MoDL in PSNR and NMSE, while providing stronger structural similarity than diffusion baselines, especially at higher acceleration.

Qualitative examples in Fig.~\ref{fig:acceleration_qualitative} show residual aliasing in zero-filled images, smoothing in UNet-like baselines, and variable local error in diffusion outputs. Yellow and blue bounding boxes from fastMRI+ indicate cartilage partial-thickness loss and meniscus tear regions, respectively. SA-RDM-DC preserved sharper anatomy and lower highlighted-region error in these pathology-annotated regions, particularly at $R=8$ and $R=12$.

\begin{figure}[!t]
\centerline{\includegraphics[width=0.95\columnwidth]{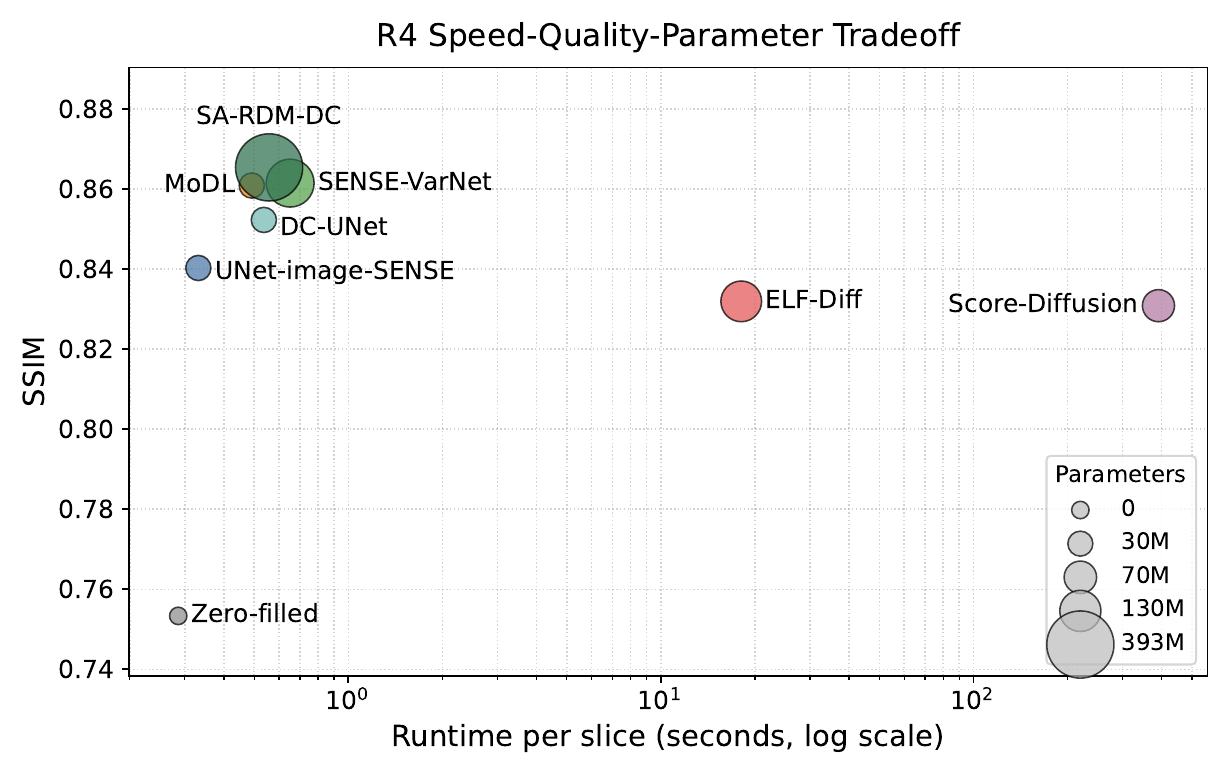}}
\caption{R4 speed-quality-parameter tradeoff. Runtime is shown on a log scale and marker size indicates model size. }
\label{fig:speed_quality_tradeoff}
\end{figure}

In the $R=4$ runtime benchmark (Fig.~\ref{fig:speed_quality_tradeoff}), SA-RDM-DC achieved the highest SSIM while requiring 0.57 s per slice, comparable to VarNet (0.65 s) and MoDL (0.49 s) and much faster than ELF-Diff (18.1 s) and Score-Diffusion (390.7 s). These results show that SA-RDM-DC combines fast feed-forward inference with strong structural image quality.

\begin{figure*}[!t]
\centerline{\includegraphics[width=0.95\textwidth]{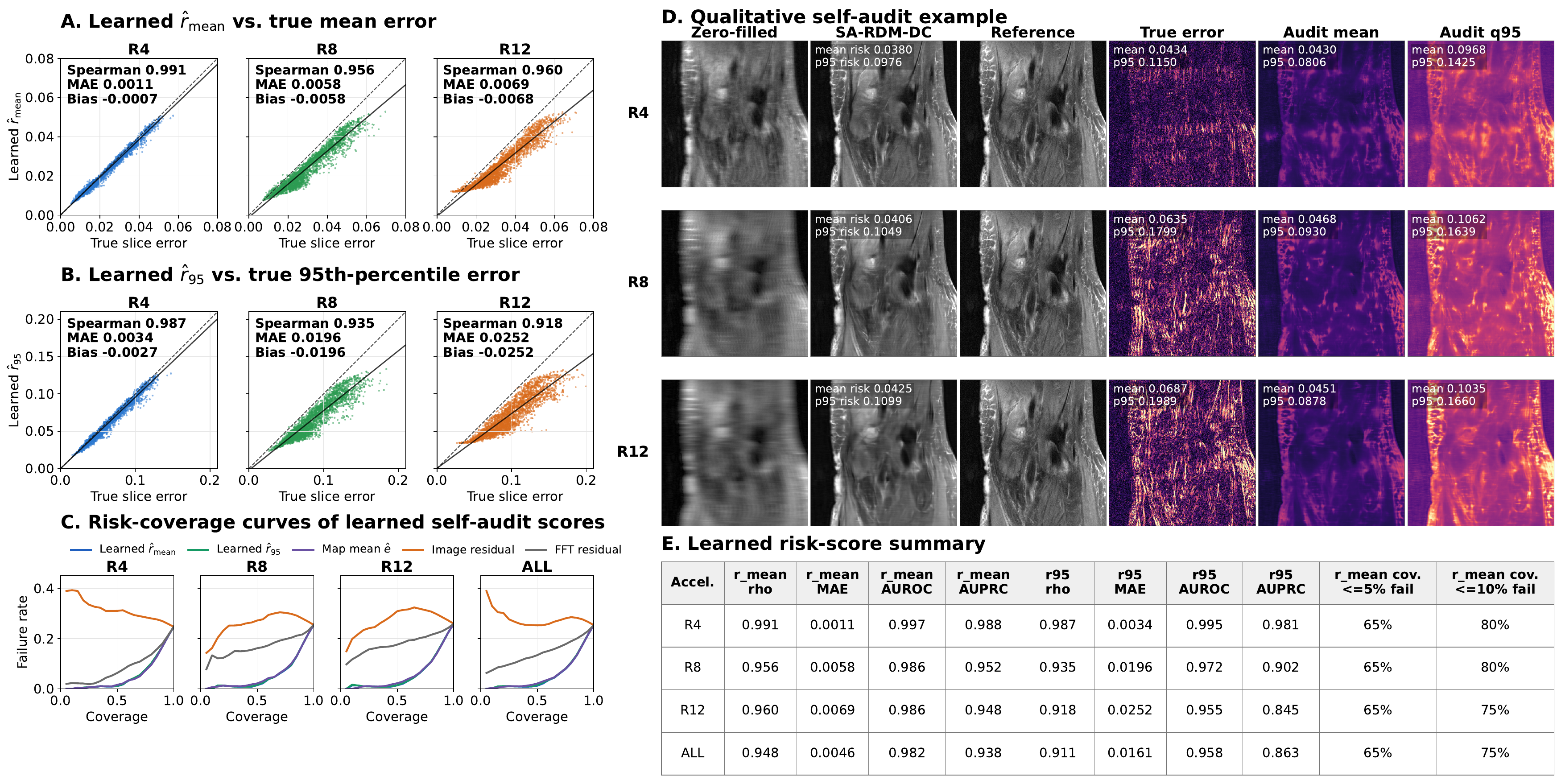}}
\caption{Self-auditing results of SA-RDM-DC on fastMRI. (A) Learned mean-risk score $\hat{r}_{\mathrm{mean}}$ versus true slice-wise mean reconstruction error at $R=4$, $R=8$, and $R=12$. (B) Learned tail-risk score $\hat{r}_{95}$ versus true slice-wise 95th-percentile reconstruction error. (C) Risk-coverage curves compare learned risk scores with image-residual and FFT-residual baselines; lower failure rate at a given coverage indicates better identification of reliable slices. (D) Qualitative self-audit examples show zero-filled input, SA-RDM-DC reconstruction, reference image, true error map, predicted audit mean map, and predicted 95th-percentile audit map across acceleration factors. (E) Summary table reports risk-score correlation, calibration error, high-error detection AUROC/AUPRC, and maximum retained coverage under 5\% and 10\% failure-rate constraints.}
\label{fig:self_auditing_results}
\end{figure*}

\subsection{Self-Auditing Reliability of SA-RDM-DC}
\label{subsec:results_self_audit}
The self-auditing head strongly tracked true reconstruction error (Fig.~\ref{fig:self_auditing_results}A-B). Spearman correlations between learned mean risk and true mean absolute error were 0.991, 0.956, and 0.960 at $R=4$, $R=8$, and $R=12$, with high-error AUROC values of 0.997, 0.986, and 0.986. The 95th-percentile risk score similarly tracked tail error, with correlations of 0.987, 0.935, and 0.918.

Risk-coverage curves in Fig.~\ref{fig:self_auditing_results}C showed that the mean audit score retained 80\% coverage at a failure rate below 10\% for $R=4$ and $R=8$, and 75\% coverage for $R=12$ and pooled accelerations. Simple image-residual scores did not retain useful coverage under the same criterion.

The qualitative audit examples in Fig.~\ref{fig:self_auditing_results}D further show that PC-SAN localized regions of elevated reconstruction error rather than simply marking all high-intensity anatomy as uncertain. This behavior is consistent with the use of both image-residual and measured-data residual features in the audit input. The mean-risk score was most useful for identifying globally poor reconstructions, whereas the tail-risk score emphasized localized high-error slices. Together, these outputs support selective review at both the slice and regional levels.

Fig.~\ref{fig:self_auditing_results}E summarizes slice-level reliability. Across accelerations, SA-RDM-DC showed high rank correlation, low calibration error, and strong high-error detection, with $r_{\mathrm{mean}}$ AUROC values of 0.997, 0.986, and 0.986.

\subsection{Pathology-Region Fidelity and Task Preservation}
\label{subsec:results_pathology}
Pathology-region analysis using fastMRI+ annotations is summarized in Table~\ref{tab:pathology_results}. SA-RDM-DC achieved the highest ROI SSIM at all acceleration factors (0.948, 0.910, and 0.877 at $R=4$, $R=8$, and $R=12$), the lowest HFEN at $R=4$ and $R=12$, and competitive edge error. In frozen classifier analysis, it had the lowest meniscus flip rate at $R=4$ (0.051) and $R=8$ (0.071), and tied VarNet at $R=12$ (0.102). Cartilage results were more variable, although SA-RDM-DC had the lowest flip rate at $R=8$.

\subsection{Ablation Study}
\label{subsec:results_ablation}
A component ablation at $R=4$ is shown in Table~\ref{tab:ablation_results}. Removing measured-k-space data consistency produced the largest degradation, reducing PSNR by 0.438 dB and SSIM by 0.0037. Removing the physics residual or missing-k-space branch caused smaller but consistent losses, indicating complementary benefits for detail recovery.

\begin{table}[!t]
\caption{R4 ablation study for SA-RDM-DC.}
\label{tab:ablation_results}
\centering
\footnotesize
\setlength{\tabcolsep}{2pt}
\begin{tabular*}{\columnwidth}{@{\extracolsep{\fill}}p{0.36\columnwidth}ccccc}
\hline
Variant & PSNR & SSIM & NMSE & MAE & HFEN \\
\hline
Full model & 34.29 & 0.8664 & 0.00976 & 0.02359 & 0.3629 \\
w/o physics residual $g_{\mathrm{zf}}$ & 33.96 & 0.8636 & 0.00994 & 0.02405 & 0.3703 \\
w/o data consistency & 33.85 & 0.8627 & 0.01008 & 0.02427 & 0.3754 \\
w/o missing-k-space branch & 34.05 & 0.8652 & 0.00981 & 0.02389 & 0.3649 \\
\hline
\end{tabular*}
\end{table}

\begin{table*}[!t]
\caption{Protocol-shift evaluation on SKM-TEA at $R=4$ under zero-shot transfer and SKM-TEA fine-tuning. mean$\pm$std.}
\label{tab:skmtea_results}
\centering
\footnotesize
\setlength{\tabcolsep}{3pt}
\begin{tabular*}{\textwidth}{@{\extracolsep{\fill}}lccc | ccc}
\hline
 & \multicolumn{3}{c}{Zero-shot} & \multicolumn{3}{c}{Fine-tuned} \\
\cline{2-7}
Method & SSIM & PSNR & NMSE & SSIM & PSNR & NMSE \\
\hline
UNet-image-SENSE & $0.6533 \pm 0.0418$ & $23.02 \pm 0.95$ & $0.1030 \pm 0.0257$ & $0.7432 \pm 0.0348$ & $25.44 \pm 1.05$ & $0.0616 \pm 0.0249$ \\
DC-UNet & $0.6900 \pm 0.0441$ & $24.01 \pm 1.09$ & $0.0824 \pm 0.0222$ & $0.7916 \pm 0.0357$ & $26.62 \pm 1.21$ & $0.0466 \pm 0.0165$ \\
SENSE-VarNet & $0.6930 \pm 0.0427$ & $24.05 \pm 1.07$ & $0.0811 \pm 0.0222$ & $0.8263 \pm 0.0347$ & $27.95 \pm 1.37$ & $0.0346 \pm 0.0128$ \\
Score-Diffusion & $0.6570 \pm 0.0473$ & $23.19 \pm 1.06$ & $0.0980 \pm 0.0232$ & $0.6562 \pm 0.0474$ & $23.17 \pm 1.06$ & $0.0984 \pm 0.0232$ \\
ELF-Diff & $0.7301 \pm 0.0478$ & $24.94 \pm 1.30$ & $0.0667 \pm 0.0207$ & $0.7323 \pm 0.0478$ & $25.00 \pm 1.31$ & $0.0659 \pm 0.0206$ \\
MoDL & $0.7006 \pm 0.0475$ & $24.36 \pm 1.14$ & $0.1051 \pm 0.0992$ & $0.8839 \pm 0.0352$ & $30.00 \pm 1.92$ & $0.0247 \pm 0.0203$ \\
SA-RDM-DC & $0.7243 \pm 0.0430$ & $24.80 \pm 1.18$ & $0.0684 \pm 0.0197$ & $0.8263 \pm 0.0315$ & $27.83 \pm 1.26$ & $0.0340 \pm 0.0092$ \\
\hline
\end{tabular*}
\end{table*}

\begin{figure*}[!t]
\centerline{\includegraphics[width=0.95\textwidth]{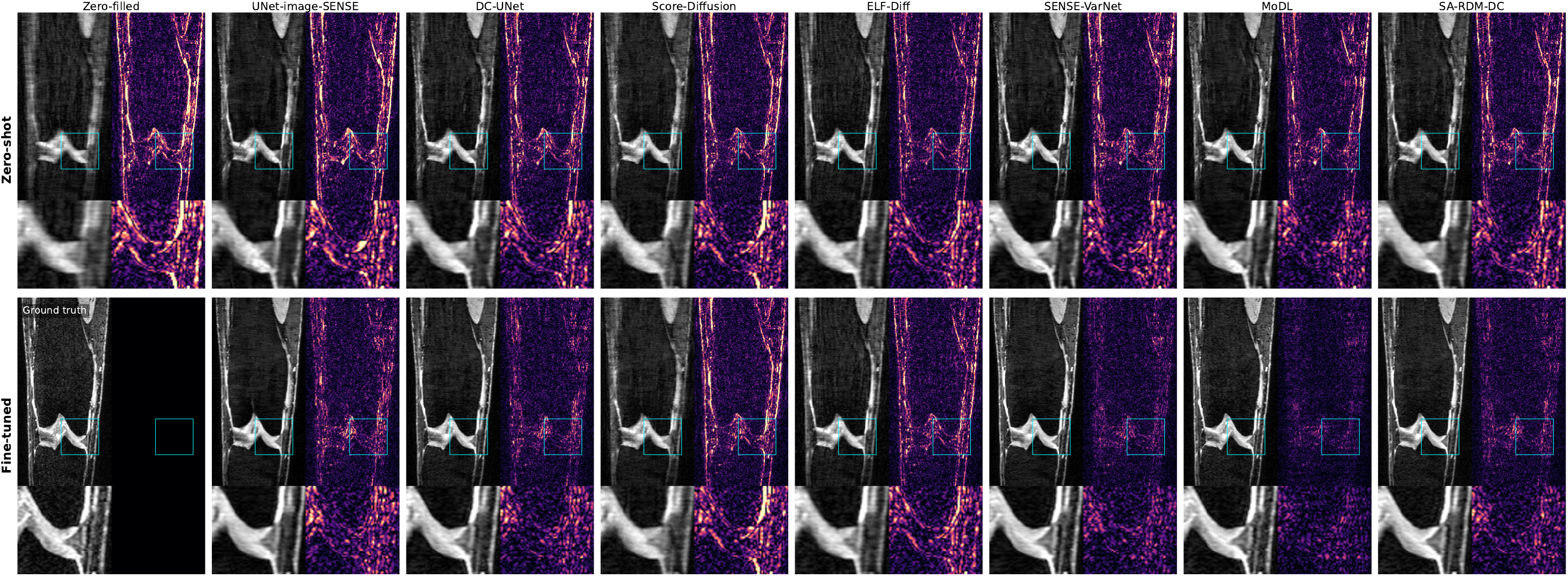}}
\caption{Protocol-shift reconstruction comparison on SKM-TEA at $R=4$. Zero-shot and fine-tuned results are shown with error maps, illustrating the effect of dataset adaptation on reconstruction quality.}
\label{fig:skmtea_reconstruction}
\end{figure*}

\subsection{Protocol-Shift Evaluation on SKM-TEA}
\label{subsec:results_skmtea}
Models trained on fastMRI were evaluated on SKM-TEA at $R=4$ in two protocol-shift settings: zero-shot transfer and target-protocol fine-tuning. Zero-shot transfer tests direct cross-protocol generalization, whereas fine-tuning tests adaptability after exposure to SKM-TEA training scans; therefore, fine-tuned performance should be interpreted as target-protocol adaptation rather than independent external validation. In zero-shot testing, SA-RDM-DC achieved SSIM 0.7243 and PSNR 24.80 dB, close to ELF-Diff and higher than VarNet and MoDL in SSIM (Table~\ref{tab:skmtea_results}). After fine-tuning, SA-RDM-DC improved to SSIM 0.8263 and PSNR 27.83 dB, matching VarNet in SSIM and exceeding VarNet in NMSE, although MoDL achieved the strongest fine-tuned global metrics. Thus, the main value of SA-RDM-DC in this setting is not global-metric dominance, but competitive reconstruction with an explicit audit signal for selective review under protocol shift.

\begin{figure}[!t]
\centerline{\includegraphics[width=0.95\columnwidth]{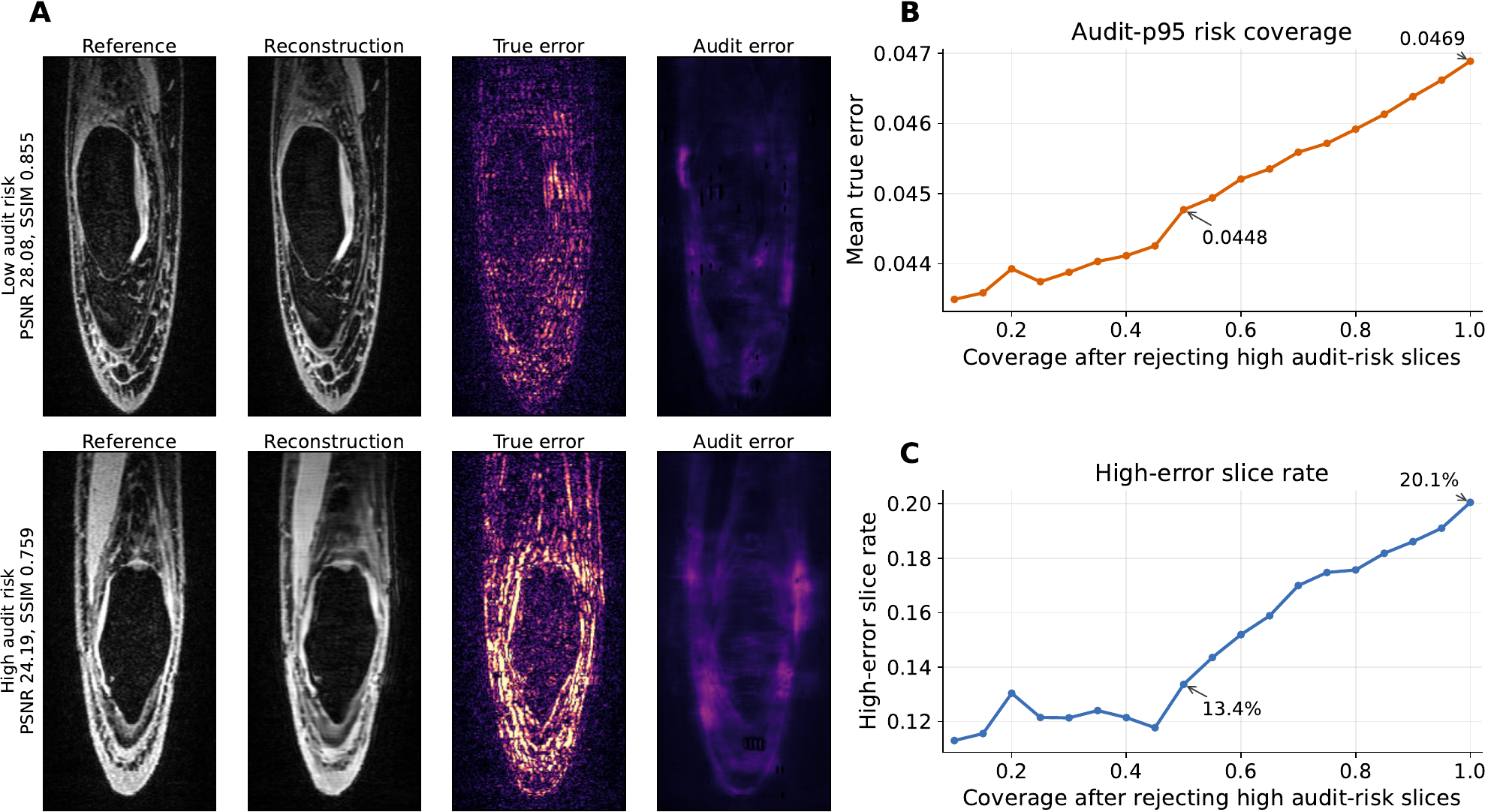}}
\caption{Self-auditing results on SKM-TEA after fine-tuning. The figure shows qualitative audit examples and risk-coverage behavior, highlighting the effect of protocol shift on audit calibration.}
\label{fig:skmtea_self_audit}
\end{figure}

Qualitative SKM-TEA examples (Fig.~\ref{fig:skmtea_reconstruction}) show visible zero-shot domain-shift effects and improved tissue boundaries after fine-tuning. SKM-TEA self-auditing was less calibrated than fastMRI (Fig.~\ref{fig:skmtea_self_audit}), but selecting the lowest-risk half of slices reduced the high-error rate from 20.1\% to 13.4\%.

\section{Discussion}
\label{sec:discussion}

This study introduced SA-RDM-DC, a self-auditing residual generative drifting model with measured-data consistency for accelerated knee MRI. It reconstructs undersampled multi-coil MRI while estimating reliability. Across fastMRI accelerations, SA-RDM-DC achieved the highest SSIM, competitive global error metrics, improved pathology-region fidelity, subsecond inference, and strong self-auditing performance, supporting evaluation beyond global fidelity to include structural preservation, pathology-region behavior, runtime, and case-specific reliability. Unlike iterative diffusion reconstructions that sample an image distribution through many reverse steps, SA-RDM-DC uses the generative drifting principle to learn a directed residual correction field for one-pass MRI reconstruction, making the generative component compatible with subsecond inference and explicit measured-data consistency.

Compared with existing MRI acceleration methods, SA-RDM-DC showed its main advantage in structural fidelity. In the global comparison in Table~\ref{tab:global_comparison}, SA-RDM-DC achieved the highest SSIM at $R=4$, $R=8$, and $R=12$, while remaining close to MoDL and SENSE-VarNet in PSNR and NMSE. MoDL achieved slightly better PSNR or error metrics in some settings, which is expected for a highly constrained unrolled model, but SA-RDM-DC provided stronger structural similarity across accelerations. This distinction is important because PSNR and NMSE do not always reflect preservation of small cartilage and meniscal structures that are clinically relevant in knee MRI.

Qualitative comparisons showed residual aliasing in zero-filled reconstructions, smoothing in U-Net-like baselines, and variable local error in iterative diffusion outputs. SA-RDM-DC better preserved boundaries and reduced highlighted-region errors, especially at higher acceleration. Pathology-region results (Table~\ref{tab:pathology_results}) showed the highest ROI SSIM across accelerations, indicating that gains extended into annotated abnormality regions rather than only background tissue. The classifier-preservation analysis showed fewer meniscus prediction flips after SA-RDM-DC reconstruction, suggesting improved stability for small structures vulnerable to blur and residual artifacts. Cartilage flip rates were less consistently improved, which may reflect broader cartilage morphology, endpoint heterogeneity, and the limited strength of the frozen classifier for this task.

Runtime further distinguishes SA-RDM-DC from iterative generative baselines. Diffusion and bridge-based methods required tens to hundreds of seconds per slice, whereas SA-RDM-DC produced reconstruction and audit outputs in one feed-forward pass, close to deterministic unrolled baselines and with additional reliability outputs.

The ablation study supports the proposed architecture. Hard measured-k-space data consistency was the dominant safeguard, as removing it degraded every reported metric and increased HFEN. The zero-filled physics residual supplied information about measured-data mismatch, while the missing-k-space branch improved learned correction in unacquired frequencies. These complementary effects support the design choice of combining image-domain residual drifting with acquisition-domain constraints, rather than relying on a purely image-domain generator or a purely deterministic data-consistency module.

The SKM-TEA experiment evaluated protocol shift to a different knee MRI sequence and voxel resolution. Zero-shot performance reflects cross-protocol generalization, whereas fine-tuned performance reflects adaptability under target-domain supervision. Fine-tuning substantially improved SA-RDM-DC, although MoDL achieved the strongest fine-tuned global metrics, suggesting that highly constrained unrolled models can be advantageous when target-protocol training data are available. The value of SA-RDM-DC under this shift is therefore competitive reconstruction plus an explicit selective-review signal. SKM-TEA audit calibration weakened compared with fastMRI, indicating that self-auditing scores should be validated or recalibrated for new protocols; nevertheless, low-risk SKM-TEA slices had lower high-error rates, suggesting partial ranking value under domain shift.

Several limitations should be noted. The primary evaluation focused on knee MRI, and generalization to other anatomies, vendors, field strengths, sequences, and sampling patterns remains untested. The SKM-TEA analysis included both zero-shot testing and target-protocol fine-tuning, so only the zero-shot setting should be interpreted as independent cross-protocol generalization. Pathology-aware analysis used fastMRI+ boxes and frozen classifier outputs rather than prospective radiologist reader studies, and the ablation study emphasized $R=4$. Future work should test multi-institutional cohorts, broader acquisition protocols, segmentation-derived quantitative biomarkers, reader-study endpoints, calibration transfer for PC-SAN, and adaptation strategies that reduce the need for target-domain fine-tuning.



\begin{thebibliography}{10}
\providecommand{\url}[1]{#1}
\csname url@samestyle\endcsname
\providecommand{\newblock}{\relax}
\providecommand{\bibinfo}[2]{#2}
\providecommand{\BIBentrySTDinterwordspacing}{\spaceskip=0pt\relax}
\providecommand{\BIBentryALTinterwordstretchfactor}{4}
\providecommand{\BIBentryALTinterwordspacing}{\spaceskip=\fontdimen2\font plus
\BIBentryALTinterwordstretchfactor\fontdimen3\font minus \fontdimen4\font\relax}
\providecommand{\BIBforeignlanguage}[2]{{%
\expandafter\ifx\csname l@#1\endcsname\relax
\typeout{** WARNING: IEEEtran.bst: No hyphenation pattern has been}%
\typeout{** loaded for the language `#1'. Using the pattern for}%
\typeout{** the default language instead.}%
\else
\language=\csname l@#1\endcsname
\fi
#2}}
\providecommand{\BIBdecl}{\relax}
\BIBdecl

\bibitem{lustig2007sparse}
M.~Lustig, D.~Donoho, and J.~M. Pauly, ``Sparse {MRI}: The application of compressed sensing for rapid {MR} imaging,'' \emph{Magnetic Resonance in Medicine}, vol.~58, no.~6, pp. 1182--1195, 2007.

\bibitem{zbontar2018fastmri}
J.~Zbontar \emph{et~al.}, ``{fastMRI}: An open dataset and benchmarks for accelerated {MRI},'' \emph{arXiv preprint arXiv:1811.08839}, 2018.

\bibitem{pruessmann1999sense}
K.~P. Pruessmann, M.~Weiger, M.~B. Scheidegger, and P.~Boesiger, ``{SENSE}: Sensitivity encoding for fast {MRI},'' \emph{Magnetic Resonance in Medicine}, vol.~42, no.~5, pp. 952--962, 1999.

\bibitem{griswold2002grappa}
M.~A. Griswold \emph{et~al.}, ``Generalized autocalibrating partially parallel acquisitions ({GRAPPA}),'' \emph{Magnetic Resonance in Medicine}, vol.~47, no.~6, pp. 1202--1210, 2002.

\bibitem{ronneberger2015unet}
O.~Ronneberger, P.~Fischer, and T.~Brox, ``{U-Net}: Convolutional networks for biomedical image segmentation,'' in \emph{Medical Image Computing and Computer-Assisted Intervention -- MICCAI 2015}, ser. Lecture Notes in Computer Science, vol. 9351.\hskip 1em plus 0.5em minus 0.4em\relax Springer, 2015, pp. 234--241.

\bibitem{schlemper2018deepcascade}
J.~Schlemper, J.~Caballero, J.~V. Hajnal, A.~N. Price, and D.~Rueckert, ``A deep cascade of convolutional neural networks for dynamic {MR} image reconstruction,'' \emph{IEEE Transactions on Medical Imaging}, vol.~37, no.~2, pp. 491--503, 2018.

\bibitem{aggarwal2019modl}
H.~K. Aggarwal, M.~P. Mani, and M.~Jacob, ``{MoDL}: Model-based deep learning architecture for inverse problems,'' \emph{IEEE Transactions on Medical Imaging}, vol.~38, no.~2, pp. 394--405, 2019.

\bibitem{hammernik2018varnet}
K.~Hammernik \emph{et~al.}, ``Learning a variational network for reconstruction of accelerated {MRI} data,'' \emph{Magnetic Resonance in Medicine}, vol.~79, no.~6, pp. 3055--3071, 2018.

\bibitem{sriram2020e2evarnet}
A.~Sriram \emph{et~al.}, ``End-to-end variational networks for accelerated {MRI} reconstruction,'' in \emph{Medical Image Computing and Computer Assisted Intervention -- MICCAI 2020}, ser. Lecture Notes in Computer Science, vol. 12262.\hskip 1em plus 0.5em minus 0.4em\relax Springer, 2020, pp. 64--73.

\bibitem{zhao2022fastmriplus}
R.~Zhao \emph{et~al.}, ``{fastMRI+}, clinical pathology annotations for knee and brain fully sampled magnetic resonance imaging data,'' \emph{Scientific Data}, vol.~9, no.~1, p. 152, 2022.

\bibitem{shaw2021quality}
R.~Shaw, C.~H. Sudre, S.~Ourselin, and M.~J. Cardoso, ``Estimating {MRI} image quality via image reconstruction uncertainty,'' \emph{arXiv preprint arXiv:2106.10992}, 2021.

\bibitem{desai2021skmtea}
\BIBentryALTinterwordspacing
A.~D. Desai \emph{et~al.}, ``{SKM-TEA}: A dataset for accelerated {MRI} reconstruction with dense image labels for quantitative clinical evaluation,'' in \emph{Proceedings of the Neural Information Processing Systems Track on Datasets and Benchmarks}, 2021. [Online]. Available: \url{https://datasets-benchmarks-proceedings.neurips.cc/paper/2021/hash/03c6b06952c750899bb03d998e631860-Abstract-round2.html}
\BIBentrySTDinterwordspacing

\bibitem{chung2022scoremri}
H.~Chung and J.~C. Ye, ``Score-based diffusion models for accelerated {MRI},'' \emph{Medical Image Analysis}, vol.~80, p. 102479, 2022.

\bibitem{cao2024hfs}
C.~Cao \emph{et~al.}, ``High-frequency space diffusion model for accelerated {MRI},'' \emph{IEEE Transactions on Medical Imaging}, vol.~43, no.~5, pp. 1853--1865, 2024.

\bibitem{shin2025elfdiff}
Y.~Shin, G.~Son, D.~Hwang, and T.~Eo, ``Ensemble and low-frequency mixing with diffusion models for accelerated {MRI} reconstruction,'' \emph{Medical Image Analysis}, vol. 101, p. 103477, 2025.

\bibitem{cui2025spiritdiffusion}
Z.-X. Cui \emph{et~al.}, ``{SPIRiT-Diffusion}: Self-consistency driven diffusion model for accelerated {MRI},'' \emph{IEEE Transactions on Medical Imaging}, vol.~44, pp. 1019--1031, 2025.

\bibitem{mirza2026fdb}
M.~U. Mirza \emph{et~al.}, ``Learning {F}ourier-constrained diffusion bridges for {MRI} reconstruction,'' \emph{IEEE Transactions on Medical Imaging}, 2026, early access.

\bibitem{deng2026drifting}
\BIBentryALTinterwordspacing
M.~Deng, H.~Li, T.~Li, Y.~Du, and K.~He, ``Generative modeling via drifting,'' \emph{arXiv preprint arXiv:2602.04770}, 2026. [Online]. Available: \url{https://arxiv.org/abs/2602.04770}
\BIBentrySTDinterwordspacing

\bibitem{wang2026rddm}
\BIBentryALTinterwordspacing
J.~Wang, Q.~Lyu, and G.~Wang, ``{RDDM}: A residual-driven drifting model for high-fidelity low-dose {CT} denoising,'' \emph{arXiv preprint arXiv:2605.17188}, 2026. [Online]. Available: \url{https://arxiv.org/abs/2605.17188}
\BIBentrySTDinterwordspacing

\bibitem{edupuganti2021uncertainty}
V.~Edupuganti, M.~Mardani, S.~Vasanawala, and J.~Pauly, ``Uncertainty quantification in deep {MRI} reconstruction,'' \emph{IEEE Transactions on Medical Imaging}, vol.~40, no.~1, pp. 239--250, 2021.

\bibitem{hu2021predict}
S.~Hu, N.~Pezzotti, and M.~Welling, ``Learning to predict error for {MRI} reconstruction,'' in \emph{Medical Image Computing and Computer Assisted Intervention -- MICCAI 2021}, ser. Lecture Notes in Computer Science, vol. 12903.\hskip 1em plus 0.5em minus 0.4em\relax Springer, 2021, pp. 604--613.

\bibitem{ekanayake2025pixcue}
M.~Ekanayake, K.~Pawar, Z.~Chen, G.~Egan, and Z.~Chen, ``{PixCUE}: Joint uncertainty estimation and image reconstruction in {MRI} using deep pixel classification,'' \emph{Journal of Imaging Informatics in Medicine}, vol.~38, no.~4, pp. 2071--2084, 2025.

\bibitem{robinson2019automated}
R.~Robinson \emph{et~al.}, ``Automated quality control in image segmentation: Application to the {UK Biobank} cardiovascular magnetic resonance imaging study,'' \emph{Journal of Cardiovascular Magnetic Resonance}, vol.~21, no.~1, p.~18, 2019.

\bibitem{mehrtash2020calibration}
A.~Mehrtash, W.~M. Wells, C.~M. Tempany, P.~Abolmaesumi, and T.~Kapur, ``Confidence calibration and predictive uncertainty estimation for deep medical image segmentation,'' \emph{IEEE Transactions on Medical Imaging}, vol.~39, no.~12, pp. 3868--3878, 2020.

\bibitem{fournel2021misaqc}
J.~Fournel \emph{et~al.}, ``Medical image segmentation automatic quality control: A multi-dimensional approach,'' \emph{Medical Image Analysis}, vol.~74, p. 102213, 2021.

\bibitem{zenk2025failure}
M.~Zenk \emph{et~al.}, ``Comparative benchmarking of failure detection methods in medical image segmentation: Unveiling the role of confidence aggregation,'' \emph{Medical Image Analysis}, vol. 101, p. 103392, 2025.

\bibitem{wang2004image}
Z.~Wang, A.~C. Bovik, H.~R. Sheikh, and E.~P. Simoncelli, ``Image quality assessment: From error visibility to structural similarity,'' \emph{IEEE Transactions on Image Processing}, vol.~13, no.~4, pp. 600--612, 2004.

\bibitem{hu2018squeeze}
J.~Hu, L.~Shen, S.~Albanie, G.~Sun, and E.~Wu, ``Squeeze-and-excitation networks,'' in \emph{Proceedings of the IEEE Conference on Computer Vision and Pattern Recognition}, 2018, pp. 7132--7141.

\bibitem{huber1964robust}
P.~J. Huber, ``Robust estimation of a location parameter,'' \emph{The Annals of Mathematical Statistics}, vol.~35, no.~1, pp. 73--101, 1964.

\bibitem{koenker1978regression}
R.~Koenker and G.~Bassett, Jr., ``Regression quantiles,'' \emph{Econometrica}, vol.~46, no.~1, pp. 33--50, 1978.

\bibitem{ilse2018attention}
M.~Ilse, J.~M. Tomczak, and M.~Welling, ``Attention-based deep multiple instance learning,'' in \emph{Proceedings of the 35th International Conference on Machine Learning}, ser. Proceedings of Machine Learning Research, vol.~80.\hskip 1em plus 0.5em minus 0.4em\relax PMLR, 2018, pp. 2127--2136.

\bibitem{geifman2017selective}
Y.~Geifman and R.~El-Yaniv, ``Selective classification for deep neural networks,'' \emph{arXiv preprint arXiv:1705.08500}, 2017.

\bibitem{loshchilov2019decoupled}
\BIBentryALTinterwordspacing
I.~Loshchilov and F.~Hutter, ``Decoupled weight decay regularization,'' in \emph{International Conference on Learning Representations}, 2019. [Online]. Available: \url{https://openreview.net/forum?id=Bkg6RiCqY7}
\BIBentrySTDinterwordspacing

\bibitem{paszke2019pytorch}
A.~Paszke \emph{et~al.}, ``{PyTorch}: An imperative style, high-performance deep learning library,'' in \emph{Advances in Neural Information Processing Systems}, vol.~32, 2019, pp. 8024--8035.

\end{thebibliography}
\end{document}